\documentclass[a4paper,11pt]{article}
\pdfoutput=1 

\usepackage{jcappub} 

\usepackage[T1]{fontenc} 

\usepackage{amssymb, amsmath, epsfig, natbib}

\renewcommand{\vec}[1]{\mathbf{#1}}
\newcommand{\vr}{\vec{r}}
\newcommand{\vq}{\vec{q}}
\newcommand{\vx}{\vec{x}}
\newcommand{\vk}{\vec{k}}
\newcommand{\vPsi}{\vec{\Psi}}

\author[a,b]{Zvonimir Vlah}
\author[c,d]{Martin White}
\author[d]{Alejandro Aviles}

\affiliation[a]{Stanford Institute for Theoretical Physics and Department of Physics, Stanford University, Stanford, CA 94306, USA} 
\affiliation[b]{Kavli Institute for Particle Astrophysics and Cosmology, SLAC and Stanford University, Menlo Park, CA 94025, USA}
\affiliation[c]{Department of Physics, University of California,
Berkeley, CA 94720}
\affiliation[d]{Department of Astronomy, University of California,
Berkeley, CA 94720}

\emailAdd{zvlah@stanford.edu}
\emailAdd{mwhite@berkeley.edu}
\emailAdd{aviles@berkeley.edu}

\title{A Lagrangian effective field theory}

\keywords{cosmological parameters from LSS -- power spectrum --
baryon acoustic oscillations -- galaxy clustering}

\abstract{
We have continued the development of Lagrangian, cosmological perturbation
theory for the low-order correlators of the matter density field.  We
provide a new route to understanding how the effective field theory (EFT) of
large-scale structure can be formulated in the Lagrandian framework and
a new resummation scheme, comparing our results to earlier work and to a
series of high-resolution N-body simulations in both Fourier and configuration
space.  The `new' terms arising from EFT serve to tame the dependence of
perturbation theory on small-scale physics and improve agreement with
simulations (though with an additional free parameter).
We find that all of our models fare well on scales larger than about two to
three times the non-linear scale, but fail as the non-linear scale is
approached.  This is slightly less reach than has been seen previously.
At low redshift the Lagrangian model fares as well as EFT in its Eulerian
formulation, but at higher $z$ the Eulerian EFT fits the data to smaller
scales than resummed, Lagrangian EFT.
All the perturbative models fare better than linear theory.}
\arxivnumber{YYMM.NNNNN}

\begin{document}
\maketitle
\flushbottom

\section{Introduction}

The Universe we observe contains structure on essentially all scales, which
is believed to arise from a process of gravitational instability acting
on small perturbations laid down in the very early Universe.
The theory of the evolution of these perturbations in their linear phase
is now well developed, and the exceptional agreement between theory and
observation for the anisotropies in the cosmic microwave background is one
of the triumphs of modern cosmology \cite{Planck15}.
Studying these evolution of these perturbations in the modern Universe,
when linear perturbation theory is breaking down, is more difficult but
can be a powerful probe of cosmology \cite{Bla10,PDG14}.

Traditionally perturbation theories in cosmology have been formulated
primarily in `Eulerian form', wherein the matter is treated as a pressureless
fluid and a perturbative solution to the continuity, Euler and Poisson
equations is obtained (see e.g.~ref \cite{Ber02} for a review).
Zeldovich \cite{Zel70} proposed a `Lagrangian form' of perturbation theory,
in which one solves perturbatively for the displacement field between
the initial and final positions of fluid elements (or dark matter particles).
Lagrangian theories tend  to fare much better at describing the large-scale
advection of matter, and allow a simpler route to understanding redshift-space
distortions and the clustering of biased objects \cite{Mat08a,Mat08b}.
Lagrangian perturbation theory (LPT) has now been well developed in the
literature
\cite{Buc89,Mou91,BouColHivJus95,Hiv95,Bha96,TayHam96,Mat08a,Mat08b,
RamBuc12,CLPT,Whi14,VlaSelBal15,ZheFri14,Mat15}
and it has been applied to understanding the broadening and shifts of the
baryon acoustic oscillation (BAO) peak \cite{PadWhi09,McCSza12,TasZal12a},
how reconstruction removes these effects
\cite{PadWhiCoh09,NohWhiPad09,TasZal12b,Whi15},
to study higher order statistics \cite{Tas14b,MohSel14}
and as the base for a new version of the halo model \cite{SelVla15}.

One of the drawbacks of both the Eulerian and Lagrangian schemes in their
classical formulation is that they are only valid prior to shell crossing
and that they treat non-linear scales as if they were perturbative.
One method for treating these deficiencies is through effective field theory
techniques, in which `effective' equations of motion which depend only on
smoothed fields are solved
\cite{BNSZ12,CHS12,PajZal13,Man14,MerPaj14,CLP14}.
A description of a Lagrangian effective theory was presented in
\cite{PorSenZal14}, and used in \cite{SenZal15} as a technique for doing IR
resummation.  An alternative formulation specialized to the case of one
spatial dimension was presented in \cite{McQWhi15}.

In this paper we present a new derivation of the Lagrangian effective field
theory, show how it can be coupled with existing resummation schemes and
compare our results to a series of N-body simulations in both Fourier and
configuration space.
The outline of the paper is as follows.
In Section \ref{sec:background} we present some  background on LPT and
effective field theory, introducing our notation and conventions.
Section \ref{sec:EoM} shows how one can derive an effective equation of
motion for the Lagrangian displacement which can be iteratively solved and
which contains the effects of short-wavelength perturbations as a series
of corrections to the usual Lagrangian perturbation theory expansion.
The cumulants of the Lagrangian displacement are all that are needed to
derive the correlation function and power spectrum of the mass, and in
section \ref{sec:cumulants} we discuss the key cumulants and the counter
terms which are introduced by EFT.
Section \ref{sec:resum} introduces our main results, a resummed version
of 1-loop Lagrangian perturbation theory which incorporates the EFT
corrections to lowest order.  This is contrasted with the Eulerian formulation
in section \ref{sec:comparisonEFT} and to compared to N-body simulations in
section \ref{sec:sims}.
We summarize our major findings in section \ref{sec:conclusions}, while
some technical details are relegated to appendices.

\section{Background}
\label{sec:background}

We shall be primarily interested in the 2-point statistics of the fractional
density perturbation, $\delta=\rho/\bar{\rho}-1$, with the correlation
function defined as $\xi(\vr) = \langle \delta(\vx) \delta(\vx+\vr) \rangle$,
and its Fourier transform, the power spectrum $P(\vk)$, defined as
$\langle \delta(\vk) \delta(\vk') \rangle =
 (2\pi)^3 \delta_D(\vk+\vk') P(\vk)$
where angled brackets signify an ensemble average.
Here and throughout $\delta_D$ denotes the 3-dimensional Dirac delta function,
and we use the Fourier transform convention
\begin{equation}
    F(\vx) = \int \frac{d^3k}{(2\pi)^3}~ F(\vk) e^{i\vk\cdot\vx} .
\end{equation}

The Lagrangian approach to cosmological structure formation was developed
in \cite{Zel70,Buc89,Mou91,Hiv95,TayHam96,Mat08a,Mat08b,CLPT,Mat15} and
traces the trajectory of an individual fluid element through space and time.
For a fluid element located at position $\vq$ at some initial time $t_0$,
its position at subsequent times can be written in terms of the Lagrangian
displacement field $\vPsi$,
\begin{equation}
    \vx(\vq,t) = \vq + \vPsi(\vq,t) ,
\label{eqn:LtoE}
\end{equation}
where $\vPsi(\vq,t_0) = 0$.  Every element of the fluid is uniquely labeled by
$\vq$ and $\vPsi(\vq,t)$ fully specifies the evolution.  Once $\vPsi(\vq)$ is
known, the density field at any time is simply
\begin{equation}
  1+\delta(\vx)=\int d^3q\ \delta_D\big[\vx-\vq-\vPsi(\vq)\big]
  \quad\Rightarrow\quad
  \delta(\vk)=\int d^3q\ e^{i\vk\cdot\vq}\big(e^{i\vk\cdot\vPsi(\vq)}-1\big)
\label{eqn:deltadefn}
\end{equation}
The evolution of $\vPsi$ is governed by
$\partial_t^2\vPsi + 2H\partial_t\vPsi = -\nabla\Phi(\vq+\vPsi)$.
We shall work throughout in terms of conformal time, $d\eta=dt/a$, and
write $\mathcal{H}=aH$ for the conformal Hubble parameter.  The equation of
motion is thus
\begin{equation}
  \ddot{\vPsi} + \mathcal{H}\dot{\vPsi} = -\nabla\Phi(\vq+\vPsi)
\label{eqn:EoM}
\end{equation}
where overdots indicate derivatives w.r.t.~conformal time.
In section \ref{sec:EoM} we describe how we account for small-scale
structures in eq.~\ref{eqn:EoM} such that the remaining fields contain only
small, long-wavelength perturbations which are amenable to treatment via
LPT.  In LPT one finds a perturbative solution for $\vPsi$:
\begin{equation}
  \vPsi(\vq,t) = \vPsi^{(1)}(\vq,t) + \vPsi^{(2)}(\vq,t)
               + \vPsi^{(3)}(\vq,t) + \cdots .
\end{equation}
with the first order solution, linear in the density field, being the
Zel'dovich approximation \cite{Zel70}.
Higher order solutions are specified in terms of integrals of higher powers
of the linear density field \cite{ZheFri14,Mat15} (see eq.~\ref{eqn:LPT}).
To these perturbative terms are then added a series of `extra' terms, which
encapsulate the effect of the small-scale physics which is missing in the
perturbative treatment.

\section{Effective equations of motion}
\label{sec:EoM}

The dynamics of our system are specified by the equations of motion
eq.~(\ref{eqn:EoM}) for which we shall attempt a perturbative solution.
However on small scales the fluctuations are large and not amenable to
a perturbative treatment, which has led the community to investigate
effective field theory descriptions which can provide an accurate description
of the long-wavelength physics without detailed knowledge of the
short-wavelength dynamics.
Following the philosophy of effective field theory as it is normally used
in the cosmology community
\cite{BNSZ12,CHS12,PajZal13,Man14,MerPaj14,CLP14}
we shall smooth eq.~(\ref{eqn:EoM}) to remove small scales, accounting for
the small-scale physics with a series of counter terms each containing
constants we are not be able to determine from the theory itself.
As emphasized by \cite{CLP14} this method has the drawback of removing
too many small-scale terms, including those generated by two long wavelength
modes, however for the low orders of interest to us it is sufficient.
We shall also restrict ourselves to the longitudinal degrees of freedom,
since again at the orders we work the effects of vorticity can be safely
ignored.
In this section we have tried to make explicit connection with the earlier
work of ref.~\cite{PorSenZal14}, who presented an investigation of
Lagrangian EFT, sometimes adopting their notation to make the connections
most clear.

We smooth eq.~(\ref{eqn:EoM}) in $\vq$-space using a filter $W_R(\vq,\vq')$,
splitting the system into $L-$long and $S-$short wavelength modes, e.g.
\begin{equation}
  \vPsi_L(\vq)=\int d^3q'\ W_R(\vq,\vq')\vPsi(\vq') \quad , \quad
  \vPsi_S(\vq,\vq')=\vPsi(\vq')-\vPsi_L(\vq)
\end{equation}
{}from which it follows that the integral of $W_R\vPsi_S$ over $\vq'$ vanishes.
By analogy we also define $\delta_L$ as the long-wavelength component of the
density perturbation (using eq.~\ref{eqn:deltadefn} with $\vPsi_L$) and
$\Phi_L$ as the gravitational potential sourced by $\delta_L$
\begin{equation}
  \nabla^2\Phi_L= \frac{3}{2}\mathcal{H}^2\Omega_m\delta_L
\end{equation}
(this is a different definition than ref.~\cite{PorSenZal14}, who perform
 an additional expansion for the source of the Poisson equation).
The short-wavelength density and potential are then
$\Phi_S=\Phi-\Phi_L$ and $\delta_S=\delta-\delta_L$

Smoothing the equation of motion for $\vPsi$
\begin{eqnarray}
  \ddot{\vPsi}_L(\vq) + \mathcal{H}\dot{\vPsi}_L(\vq)
  &=& -\int d^3q'\ W_R(\vq,\vq')\nabla \Phi\big(\vq'+\vPsi(\vq')\big)
  \nonumber \\
  &=& -\int d^3q'\ W_R(\vq,\vq')\nabla\Phi_L\big(\vq'+\vPsi(\vq')\big)
      -\int d^3q'\ W_R(\vq,\vq')\nabla\Phi_S\big(\vq'+\vPsi(\vq')\big)
  \nonumber \\
  &=& -\nabla\Phi_L(\vq+\vPsi_L(\vq) )
      +\vec{a}_S\big(\vq,\vPsi_L(\vq) \big) .
\label{eq:eom_v1}
\end{eqnarray}
Apart from the $\Phi$ dependence on $\vPsi_L$ we shall not need explicit
expressions for the sources in what follows, since their structure will be
dictated by symmetry.  However to make contact with ref.~\cite{PorSenZal14}
we note that the second term on the r.h.s.~can be written as a multipole
expansion having contributions from $\Phi_L$ and $\Phi_S$.  The $\Phi_L$
piece is
\begin{eqnarray}
  \vec{a}_S(\vq)
  &\ni& \nabla\Phi_L(\vq+\vPsi_L(\vq))
           -\int d^3q'\ W_R(\vq,\vq')
           \nabla \Phi_L\big(\vq'+\vPsi(\vq')\big) \nonumber \\
  &=& \int\frac{d^3k}{(2\pi)^3}\ (i\vk)\Phi_L(\vk)
   e^{i\vk\cdot(\vq+\vPsi_L(\vq))}
  \int d^3q'\ \Big(W_R(\vq,\vq')e^{i\vk\cdot\vPsi_S(\vq,\vq')}
  -\delta_D(\vq-\vq')\Big)e^{i\vk\cdot\delta\vq} \nonumber \\
  &=&\sum^{\infty}_{n=2} \frac{i^n}{n!}
  \int \frac{d^3 k}{(2\pi)^3}\ (i\vk\ k_{i_1}\ldots k_{i_n})
  \Phi_L(\vk)e^{i\vk\cdot(\vq+\vPsi_L(\vq))} \nonumber \\
  &\times&
  \int d^3q'\ W_R(\vq,\vq')
  \big[\delta\vq + \vPsi_{S}(\vq,\vq') \big]_{i_1} \ldots
  \big[\delta\vq + \vPsi_{S}(\vq,\vq') \big]_{i_n} \nonumber \\
  &=& \sum^{\infty}_{n=2} \frac{i^n}{n!}
  Q_S^{i_1\ldots i_n}(\vq)
  \int \frac{d^3k}{(2\pi)^3}\ (i\vk\ k_{i_1}\ldots k_{i_n})
  \Phi_L(\vk)e^{i\vk\cdot(\vq+\vPsi_L(\vq))} \nonumber \\
  &=& -\frac{1}{2} Q^{ij}_{S}(\vq)
  \nabla \nabla_i \nabla_j \Phi_L(\vq+\vPsi_L(\vq)) + \ldots
\end{eqnarray}
where we have defined multipole moments of the short displacement field
\begin{equation}
  Q_S^{i_1\ldots i_n}(\vq) = \int d^3q'\ W_R(\vq,\vq')
  \big[ \delta\vq + \vPsi_{S}(\vq,\vq') \big]_{i_1} \ldots
  \big[ \delta\vq + \vPsi_{S}(\vq,\vq') \big]_{i_n} .
\end{equation}
and we note that the dipole moment is missing since the averages of
$\vPsi_S$ and $\delta\vq$ both vanish.
The acceleration due to the short wavelength modes follows a similar
structure,
$\vec{a}_S(\vq) \ni -\nabla\Phi_S(\vq+\vPsi_L(\vq))
  -\frac{1}{2} Q^{ij}_{S}(\vq)
 \nabla \nabla_i \nabla_j \Phi_S(\vq+\vPsi_L(\vq)) + \ldots$,
representing the contribution of $\Phi_S$ to the evolution of $\vPsi_L$.
Note that $\Phi_S$ depends on the long-wavelength displacement (and hence
density) through its argument but there is no contribution to the center
of mass so the dependence is through tidal fields of $\Phi_L$.

Regardless of the particular form for the expansion we have
\begin{equation}
  \ddot{\vPsi}_L + \mathcal{H}\dot{\vPsi}_L = -\nabla\Phi_L(\vq+\vPsi_L(\vq))
  + \vec{a}_S \big(\vq,\vPsi_L(\vq) \big)
\end{equation}
where the first term can be treated perturbatively and the ``extra'' term,
$\vec{a}_S$, contains sources of displacement that arise from small-scale
modes which may not be well captured by perturbation theory.\footnote{In our
approach all of the short-distance terms, including the multipole expansion
of $\Phi$, are absorbed in $\vec{a}_S$ in contrast to ref.~\cite{PorSenZal14}.
Thus the `additional' terms all arise from $\vec{a}_S$.}
The contribution to $\vPsi_L$ which is $n^{\rm th}$ order in the
long-wavelength, linear theory perturbation, $\delta_0$, is
\begin{equation}
  \vPsi^{(n)}_L(\vk) = \frac{iD^n}{n!}
  \int\frac{d^3k_1}{(2\pi)^3}\cdots\frac{d^3k_n}{(2\pi)^3}
  (2\pi)^3\delta_D\big(\sum_j \vk_j-\vk\big) \vec{L}_n(\vk_1,\cdots,\vk_n)
  \delta_0(\vk_1)\cdots\delta_0(\vk_n)
\label{eqn:LPT}
\end{equation}
with $D$ the linear growth rate and the $\vec{L}_n$ given in
e.g.~\cite{Mat08a,ZheFri14,Mat15}
with the lowest order term being simply $\vec{L}_1(\vk)=\vk/k^2$.

The additional contributions to $\vPsi_L$ come from the source term,
$\vec{a}_S$, which we must integrate against the Green's function.
We cannot compute this term from first principles, but we can parameterize its
dependence on $\vPsi_L$ with a small number of terms which are constrained by
the symmetries of the problem.  The first contribution is a `stochastic' term,
$\mathcal{S}$, which is independent of the long-wavelength modes.
The first non-trivial dependence on the long-wavelength density that transforms
as a vector must\footnote{In terms of $a_S$ this term comes from taking
$Q_S^{ij}\propto\delta^{ij}$ and noting that the long-wavelength modes
contribute to $a_S$ as tidal fields.}
be proportional to $\nabla\delta_0$.
Grouping the contributions by their dependence on $\delta_0$ and the number
of spatial derivatives (see ref.~\cite{PorSenZal14,McQWhi15,Bal15} for similar
expansions) and keeping only the lowest order terms we thus have
\begin{equation}
  \vPsi_L \ni \mathcal{S} + \frac{1}{2}\alpha_1\nabla\delta_0 + \cdots
\end{equation}
with $\alpha_1$ an undetermined coefficient and $\mathcal{S}$ uncorrelated
with $\delta_0$.
This is the same, lowest order contribution as derived in
\cite{PorSenZal14,McQWhi15}.  As we shall see, these terms lead to
corrections to the displacement power spectrum and additionally serve to
tame contributions from high $k$ modes in the usual perturbative treatment.

Formally the expansion above is in powers of derivatives or $k/\Lambda$,
with $\Lambda$ some cut-off scale chosen to render the perturbation theory
integrals well behaved.
The theory is $\Lambda$-independent if all terms in the expansion are kept,
but doing so introduces an infinite number of undetermined constants.
If we truncate the expansion at a fixed order, and if $k<\Lambda$, higher order
terms are suppressed by powers of $k/\Lambda<1$ and thus should be numerically
smaller than the terms kept (unless some constants artificially make some
terms numerically large while being parametrically small).
Unfortunately, if we keep only the lowest order terms in $\vec{a}_S$, while
simultaneously cutting off the perturbation theory integrals at
$\Lambda\simeq k_{\rm nl}$, the theory depends on the cut-off $\Lambda$ unless
we work at very small $k$
(where all of the perturbation theory corrections are anyway small).
We shall follow the standard practice in the field and take the limit
$\Lambda\to\infty$ when computing the perturbation theory integrals and
keep only the lowest order contributions to $\vec{a}_S$.

\section{Cumulants}
\label{sec:cumulants}

The arguments of section \ref{sec:EoM} lead us to our expression for
the displacement:
\begin{equation}
  \vPsi(\vq) = \vPsi_L^{(1)}(\vq) + \vPsi_L^{(2)}(\vq) +
               \vPsi_L^{(3)}(\vq) + \cdots +
               \frac{1}{2}\alpha_1\nabla\delta_0 + \mathcal{S} + \cdots
\label{eqn:Psi}
\end{equation}
where the first three terms come from the perturbative treatment of
the long-wavelength evolution and the last few terms parameterize the
impact of the short-wavelength modes on the evolution.
The correlation function and power spectrum can now be defined through
the cumulants of the displacement.  Defining \cite{Mat08a,CLPT}
\begin{equation}
  K(\vq,\vk) = \left\langle e^{i\vk\cdot\vec{\Delta}}\right\rangle
  \quad {\rm with} \quad
  \vec{\Delta}(\vq) = \vPsi(\vq)-\vPsi(\vec{0})
\end{equation}
the power spectrum is
\begin{equation}
  P(k) = \int d^3q\ e^{i\vq\cdot\vk} \left[ K(\vq,\vk) - 1 \right]
\label{eqn:pofk}
\end{equation}
and the correlation function
\begin{equation}
  1+\xi(r) = \int \frac{d^3q\,d^3k}{(2\pi)^3}\ e^{i\vk\cdot(\vq-\vr)}
  K(\vq,\vk)
\label{eqn:xofr}
\end{equation}
If only terms quadratic in $k$ are kept in the exponential, the $k$-integral
in eq.~(\ref{eqn:xofr}) can be done analytically
(see e.g.~appendix \ref{app:identities}).
The expectation value of the exponential can be obtained using the
cumulant theorem \cite{Mat08a,Mat08b} so we can write
\begin{equation}
  \log K(\vq,\vk) = -\frac{1}{2}k_ik_jA_{ij}(\vq)
                    +\frac{i}{6}k_ik_jk_\ell W_{ij\ell}(\vq) + \cdots
\end{equation}
with
\begin{eqnarray}
  A_{ij}(\vq) &=& 2\left\langle\Psi_i(\vec{0})\Psi_j(\vec{0})\right\rangle
                 -2\left\langle\Psi_i(\vq_1)\Psi_j(\vq_2)\right\rangle
  \equiv 2\left( \Sigma^2\delta_{ij}-\eta_{ij}\right) \\
  W_{ij\ell}(\vq)&=&
    \left\langle\Psi_{\{i}(\vq_1)\Psi_j(\vq_2)\Psi_{\ell\}}(\vq_2)
    \right\rangle -
    \left\langle\Psi_{\{i}(\vq_2)\Psi_j(\vq_1)\Psi_{\ell\}}(\vq_1)
    \right\rangle
\end{eqnarray}
where we have written $\vq=\vq_1-\vq_2$ and followed the notation of
\cite{CLPT}.
Regular LPT can be obtained by Taylor series expanding the exponential and
collecting terms in powers of the linear theory power spectrum (which we
shall denote as $P_0$, for 0-loop).
The series expansion so produced agrees with Eulerian perturbation theory
\cite{RamBuc12,ZheFri14,McQWhi15,Mat15}.
Various useful resummation schemes can be introduced by keeping some of the
pieces exponentiated while expanding others (see section \ref{sec:resum}).

We now consider the contributions to $A_{ij}$ and $W_{ij\ell}$ arising from
the various orders in LPT and from the counter terms in eq.~(\ref{eqn:Psi}).
We shall denote the terms arising from the $1^{\rm st}$ order in LPT
(i.e.~Zeldovich approximation) as with a superscript ``lin'' as we shall
want to treat these terms separately on occasion.  The other terms we
shall denote ``lpt'' and ``eft'' depending on their source in
eq.~(\ref{eqn:Psi}).  Thus, e.g.,
\begin{equation}
  A_{ij}(\vq) = A_{ij}^{\rm lin}(\vq) + A_{ij}^{\rm lpt+eft}(\vq)
\end{equation}
In the $W_{ijk}$ term, products of two displacement fields appear evaluated
at the same point in space, so $W_{ijk}$ is a composite operator which
introduces new counter terms.
Even though these counter terms can be formally derived from the source,
$\vec{a}_S$, one can also obtain their form based on symmetry arguments
(see e.g.~ref.~\cite{PorSenZal14} for extensive discussion).
Considering all two-index quantities which depend at most linearly on
$\vPsi$ we thus have
\begin{eqnarray}
  \Psi_i(\vq)\Psi_j(\vq) &=&
  \Psi^{(1)}_i(\vq)\Psi^{(1)}_j(\vq) +
  \Psi^{(1)}_i(\vq)\Psi^{(2)}_j(\vq) +
  \Psi^{(2)}_i(\vq)\Psi^{(1)}_j(\vq) + \cdots \nonumber \\
  &+& \frac{1}{3}\alpha_0 \delta_{ij} + \bar\alpha_2\delta_{ij}
  \nabla_\ell \Psi^{(1)}_\ell + \bar\alpha_3
  \left[ \nabla_i \Psi^{(1)}_j + \nabla_j \Psi^{(1)}_i \right] + \ldots
\end{eqnarray}
If we were to restrict the perturbation theory integrals to $k<\Lambda$, the
coefficients (and their higher-order counterparts) would serve to make the
final results $\Lambda$ independent.  The $\Lambda$ dependence of these terms
is thus set by the structure of the high-$k$ sensitivity in the theory.  There
will also be a $\Lambda$-independent (or `finite') piece which can in
principle be different for each term.

The terms coming from the small-scale modes, which we shall call the ``EFT
terms'', contribute to $A_{ij}$ as simple integrals of the linear theory
power spectrum.
For example, the cross-term arising from linear theory in $\vPsi_L$ and
the $\nabla\delta$ term gives
\begin{eqnarray}
  \left\langle\Psi_i^{(1)}(\vq_1)\nabla_j\delta(\vq_2)\right\rangle
  &=& \left\langle \int \frac{d^3p_1\,d^3p_2}{(2\pi)^6}
  e^{i(\vec{p}_1\cdot\vec{q}_1+\vec{p}_2\cdot\vec{q}_2)}
  \frac{i\vec{p}_{1i}}{p_1^2}\delta_0(\vec{p}_1)
  i\vec{p}_{2j}\delta_0(\vec{p}_2) \right\rangle \\
  &=& \int \frac{d^3k}{(2\pi)^3} e^{i\vec{k}\cdot\vec{q}} P_0(k)
      \frac{k_ik_j}{k^2} \\
  &=& \frac{1}{3}\xi_0(q)\delta_{ij} -
      \xi_2(q)\left(\frac{1}{3}\delta_{ij}-\hat{q}_i\hat{q}_j\right)
\end{eqnarray}
with $\xi_\ell$ being the usual moments of the linear theory correlation
function:
\begin{equation}
  \xi_\ell(q) \equiv i^\ell \int \frac{k^2\,dk}{2\pi^2} P_0(k) j_\ell(kq)
\end{equation}

\begin{figure}
\begin{center}
\resizebox{1.02\columnwidth}{!}{\includegraphics{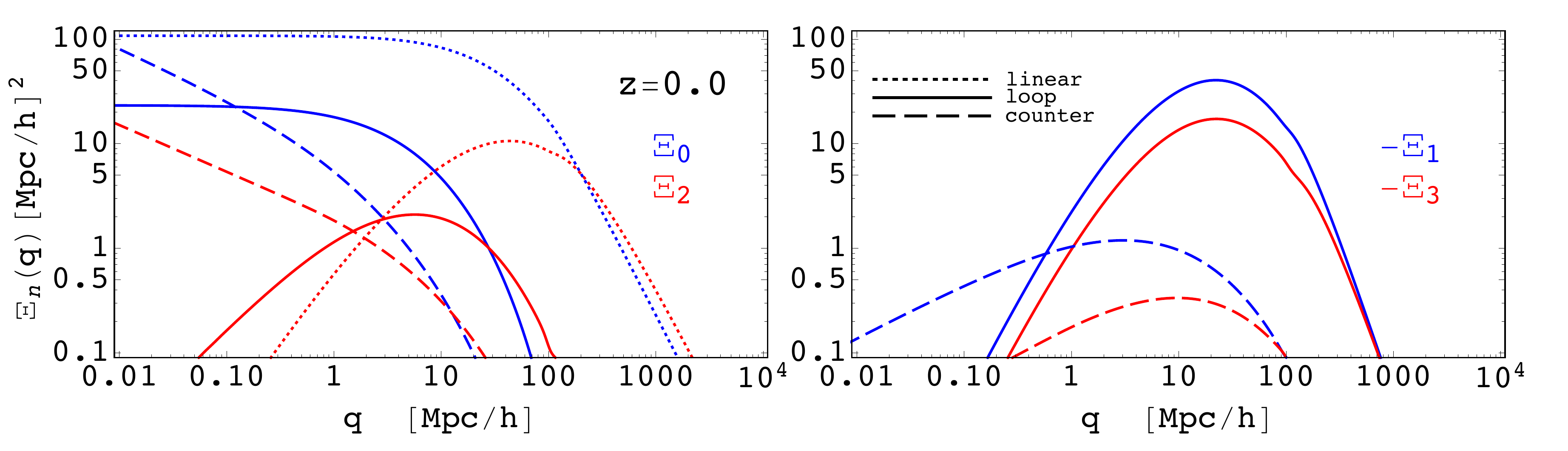}}
\end{center}
\caption{The terms entering into the cumulants of $\vPsi$, evaluated at $z=0$,
divided into contributions from the linear and 1-loop orders and the
counter-terms.  The $\Xi_\ell$ are defined in section \ref{sec:cumulants}.
Note that they are very smooth functions.
The linear pieces scale as $D^2(z)$ while the 1-loop terms scale as $D^4(z)$.
The counter terms have been plotted assuming $\alpha_n=1$.
The $\Xi_\ell$ with $\ell>0$ all have a characteristic scale.}
\label{fig:terms}
\end{figure}

The final expressions can be cast in simple form if we introduce $\Xi_\ell$
which are extensions of the $\xi_\ell$ above.
If we write
\begin{eqnarray}
  \label{eqn:En_terms}
  \Xi_{0}(q) &=& \int\frac{dk}{2\pi^2} \left[
  P_0(k) + \frac{9}{98}Q_1(k) + \frac{10}{21}R_1(k)
  + \alpha_1 k^2P_0(k)
  \right]j_0(kq)  \\
  \Xi_{1}(q) &=& \int\frac{dk}{2\pi^2}\left(-\frac{3}{7k}\right) \left[
  Q_1(k) - 3Q_2(k) + 2R_1(k) - 6R_2(k) + \alpha_2k^2P_0(k)
  \right]j_1(kq) \nonumber \\
  \Xi_{2}(q) &=& \int\frac{dk}{2\pi^2} \left[
  P_0(k) + \frac{9}{98}Q_1(k) + \frac{10}{21}R_1(k)
  + \alpha_1 k^2P_0(k)
  \right]j_2(kq) \nonumber \\
  \Xi_{3}(q) &=& \int\frac{dk}{2\pi^2}\left(-\frac{3}{7k}\right) \left[
  Q_1(k) + 2Q_2(k) + 2R_1(k) + 4R_2(k) + \alpha_3k^2P_0(k)
  \right]j_3(kq) \nonumber
\end{eqnarray}
with $Q_i$ and $R_i$ the mode-coupling integrals defined in
appendix \ref{app:identities} and $\alpha_i$ undetermined constants,
then we have
\begin{equation}
  \frac{1}{2} A_{ij}
  = \frac{1}{3}\delta_{ij}\left(\Xi_{0}(0)+\alpha_0-\Xi_{0}(q)\right)
  + \left(\hat{q}_i\hat{q}_j-\frac{1}{3}\delta_{ij}\right)\Xi_{2}(q)
\end{equation}
where $\alpha_0$ is a counter-term with contributions from the zero-lag
correlator of $\mathcal{S}$ among other places and we have dropped terms
like $\langle\mathcal{S}(\vq)\mathcal{S}(\vec{0})\rangle$.
A similar calculation gives
\begin{equation}
  W_{ij\ell}(q) = \frac{2}{5}\hat{q}_{\{i}\delta_{j\ell\}}\Xi_{1}(q)
  +\frac{3}{5}\left(5\hat{q}_i\hat{q}_j\hat{q}_\ell-
   \hat{q}_{\{i}\delta_{j\ell\}}\right)\Xi_{3}(q)
\label{eqn:A_term}
\end{equation}
and the coefficients $\alpha_2$ and $\alpha_3$ are linear combinations of the
$\bar{\alpha}_2$ and $\bar{\alpha}_3$ we introduced previously.
These terms lead to very small corrections and we shall set them to zero
henceforth.
The common term in $\Xi_0$ and $\Xi_2$ above is the displacement power
spectrum (more details on the derivation of these terms can be found in
appendix B of ref.~\cite{CLPT} or
appendix A of ref.~\cite{VlaSelBal15}).
Note that the $W_{ij\ell}$ term does not have a linear theory contribution
and so is expected to be small on large scales.
The `extra' terms, proportional to $\alpha_1$, are moments of the linear
theory correlation function.

Figure \ref{fig:terms} shows these $\Xi_\ell$ at $z=0$ as a function of the
Lagrangian displacement, $q$.  Note that they are all smooth functions
and $\Xi_\ell$ for $\ell>0$ pick out a characteristic scale.
The figure shows the contributions from linear terms, scaling as $D^2(z)$,
the 1-loop terms, scaling as $D^4(z)$, and the EFT terms.
If we integrate to finite $k$, or if $P(k)$ has a high-$k$ cut-off, the only
non-vanishing contribution as $q\to 0$ is from $\Xi_0$.

At this stage it appears that we have four different counter terms in the
1-loop predictions
(eq.~\ref{eqn:En_terms} and $\alpha_0$ from eq.~\ref{eqn:A_term}).
However, the $\alpha_0$ and $\alpha_1$ terms are the same, so these two
parameters are completely degenerate. 
The $\alpha_2$ and $\alpha_3$ terms are not identical, but on large scales
($k<0.15\,h^{-1}$Mpc) they are well approximated by the $\alpha_0$ term.
At smaller scales the differences between these terms becomes significant,
but on these scales we may expect 2-loop terms to be large and we leave the
regime of validity of our calculation. 
For these reasons we restrict ourselves henceforth to one free parameter
(we choose $\alpha_0$) and set the other three to zero. 

\section{Resummation schemes}
\label{sec:resum}

Having derived the form for the cumulants we can now use
eqs.~(\ref{eqn:pofk}, \ref{eqn:xofr}) to compute the 2-point function of
the mass, in real space.
We show in the next section that expanding the exponentials in a Taylor
series reproduces the standard, Eulerian EFT expressions to 1-loop order.
By keeping some or all of the terms in the exponential, i.e.~(re)summing
multiple orders in perturbation theory, we can derive other approximations
to the 2-point function.
As has been discussed in the literature previously
\cite{Tas14a,CLPT,Whi14,SenZal15,McQWhi15}
one particular advantage of Lagrangian approaches over Eulerian ones is
the ability to capture the main effects of advection of mass due to
long-wavelength perturbations and to sum the terms in the perturbation
theory which scale as the displacement variance times the second derivative
of the (linear) correlation function.  In $\Lambda$CDM models which contain
BAO features even at high order such terms can be numerically quite large,
despite being parametrically small.

Unfortunately we are only computing the EFT terms as a series in derivatives,
and we need to keep the number of terms small in order to limit the number
of free parameters.  These terms act to tame (or regularize) the behavior of
the perturbation theory terms, so we would like to consistently keep these
terms together.  
Perhaps the simplest way to achieve these goals is to keep in the exponential
only $A_{ij}^{\rm lin}$ while keeping only the $1^{\rm st}$ order expansion
of the other terms.  This is also numerically straightforward and efficient.
For the power spectrum, for example, we would then have
\begin{equation}
  P(k) = \int d^3q\ e^{i\vk\cdot\vq-(1/2)k_ik_jA_{ij}^{\rm lin}}
  \left[ 1 - \frac{1}{2}k_ik_jA_{ij}^{\rm lpt+eft}
           + \frac{i}{6}k_ik_jk_k W_{ijk}^{\rm lpt+eft} + \cdots \right]
\label{eqn:resum}
\end{equation}
and the correlation function can be simplified using the formulae in
appendix \ref{app:identities} as
\begin{equation}
  1+\xi(r) = \int\frac{d^3q}{(2\pi)^{3/2}|A_{\rm lin}|^{1/2}}
  e^{-(1/2)(\vr-\vq)^T \vec{A}_{\rm lin}^{-1} (\vr-\vq)}
  \left[ 1 - \frac{1}{2} G_{ij} A_{ij}^{\rm lpt+eft} +
  \frac{1}{6}\Gamma_{ijk}W_{ijk}^{\rm lpt+eft} + \cdots \right]
\end{equation}
One of the useful properties of this resummation is that the result on
large scales is quite insensitive to the details of the ``lpt+eft'' terms
on small scales, where they are not well constrained.
If terms beyond linear are to be kept in the exponent, then care must be
taken to ensure the exponentiated matrix is well-behaved and that all of
the eigenvalues are positive.
After some experimentation we found that the predictions for both $P(k)$
and $\xi(r)$ were relatively insensitive to which terms we kept exponentiated,
as long as we consistently treated the zero-lag and $q$-dependent pieces,
and we adopted the above procedure due to its numerical and algebraic
simplicity.
We shall refer to this expression as ``CLEFT'', since it is an effective
field theory in the same spirit as convolution Lagrangian perturbation
theory (CLPT; \cite{CLPT}).
The integrals can be done, numerically, by simple quadratures.
The azimuthal part of the $d^3q$ integral gives $2\pi$ so one needs to do
only the integrals over $\mu_q$ and $|\vq|$.  Alternatively one can expand
the $\mu$-dependence as in ref.~\cite{VlaSelBal15} and reduce the remaining
integral to a sum of FFTs (see Appendix \ref{app:identities} for useful
formulae).

In the above scheme both the IR-sensitive and UV-sensitive terms in
$A_{\rm lin}$ are kept exponentiated.  Other schemes have been put forth.
Motivated in part by the poor performance of `standard' EFT in describing the
broadening of the acoustic peak in the correlation function, Senatore and
Zaldarriaga put forth an IR-resummation scheme \cite{SenZal15} which is
similar in spirit to the schemes described above.
We describe explicitly the relationship between our schemes and theirs in
Appendix \ref{app:IRresum}, where we show that their scheme results in
resumming a subset of the terms in eq.~(\ref{eqn:resum}).
McQuinn \& White \cite{McQWhi15} put forth a scheme similar to ours
except that they resummed only the low-$k$ parts of $A_{\rm lin}$ while
expanding the high-$k$ pieces and the counter terms.
They found that the final results were relatively insensitive to the
split between the IR- and UV-sensitive pieces, and in fact keeping all of
the contributions exponentiated was almost identical to making the split.
For this reason we have not attempted such a split here, though it would
also be numerically straightforward.

\section{Comparison to Eulerian theory}
\label{sec:comparisonEFT}

\begin{table}
\begin{center}
\begin{tabular}{ccccc}
     & $z=0.25$ & $z=0.50$ & $z=0.75$ & $z=1.00$ \\
LEFT & $-13.6$  &  $-7.8$  & $-4.8$   & $-2.7$    \\
ZEFT & $-28.5$  & $-21.9$  & $-16.2$  & $-12.75$   
\end{tabular}
\end{center}
\caption{The adopted values of $\alpha_0$ for LEFT (eq.~\ref{eqn:resum})
and ZEFT (eq.~\ref{eqn:zeft_pk2}) in figure \ref{fig:ps}.}
\end{table}

It is instructive at this point to illustrate that our prescription regains
the standard, Eulerian result.  To obtain this we simply expand the
exponential in eq.~(\ref{eqn:pofk}) and keep terms to order $P_0^2$.
A straightforward calculation (see \cite{CLPT,VlaSelBal15} for further
details) gives
\begin{eqnarray}
  P(k) &=& P_0(k) + \frac{9}{98}Q_1(k) + \frac{3}{7}Q_2(k) + \frac{1}{2}Q_3(k)
        +  \frac{10}{21}R_1(k)+\frac{6}{7}R_2(k) - \alpha_{\text{eft}} k^2 P_0(k) \\
       &=& P_0(k) + P_{\rm 1-loop}^{SPT} - \alpha_{\text{eft}} k^2 P_0(k)
\end{eqnarray}
where we have written $\alpha_\text{eft}=\tfrac{1}{3}\alpha_0-\alpha_1-\tfrac{3}{35}(\alpha_3-\alpha_2)$ 
and indicated the equivalence with the expression for standard, 1-loop perturbation theory in the second line.
In the EFT literature this is often written
\begin{equation}
  P(k) = P_0(k) + P_{\rm 1-loop}^{SPT}
       - \left(c_s^2k_{\rm nl}^{-2}\right) k^2 P_0(k)
\end{equation}
which is obviously of the same form.
Note that the coefficient of the $k^2P_0$ term is a degenerate combination
of two different coefficients in the Lagrangian scheme.

\section{Zeldovich EFT}
\label{sec:zeft}

At this point it is interesting to consider the combination of the lowest
order terms in perturbation theory and the lowest order correction from the
small-scale terms.  Such terms are almost trivial to compute, numerically,
and could be expected to capture the essential physics at large scales.
The resulting form is also simple enough that generalizations to include
redshift-space distortions, bias and reconstruction are relatively
straightforward.

In configuration space, keeping only the linear theory LPT terms and the
$\nabla\delta$ term all of the required integrals are Gaussians and we have
\begin{equation}
  1 + \xi(r) = \int \frac{d^3q}{(2\pi)^{3/2} |A|^{1/2}}
  \ e^{-\frac{1}{2} (\vec{r}-\vec{q})^T \mathbf{A}^{-1} (\vec{r}-\vec{q})}
  \left[ 1 -  \frac{1}{3} \alpha_0\, {\rm tr}G + \cdots \right] ,
\label{eqn:xi_ZEFT}
\end{equation}
where $\mathbf{A}\equiv \mathbf{A}_{\rm lin}$ and we have used the
notation of \cite{CLPT}, reviewed in appendix \ref{app:identities}.
We shall refer to this as the ``ZEFT'' correlation function, since it is a
combination of the Zeldovich approximation and the lowest order EFT terms.
The tr$G$ term scales as a second derivative of $\xi$, and thus this
term\footnote{A similar dependence arises to leading-order as the
scale-dependent piece of peaks bias, thus this term can do double duty in
a theory of biased tracers.}
modifies the width of the BAO peak at $r\approx 110\,h^{-1}$Mpc.  It
also leads to a slight increase in the amplitude to smaller $r$.

The expression for the power spectrum is similar
\begin{eqnarray}
  P_{\rm ZEFT}(k) &=& \left(1-\tfrac{1}{3}\alpha_0 k^2\right)\int d^3q\ e^{i\vk\cdot\vq}
  e^{-\frac{1}{2}\vk^T \vec{A}\vk} \\
  &\approx& e^{-k^2\Sigma^2} \left( 1 -\tfrac{1}{3} \alpha_0 k^2 \right) P_0(k)
  \label{eqn:zeft_pk1} \\
  &\approx& e^{-k^2\Sigma^2}P_0(k) -\tfrac{1}{3} \alpha_0 k^2P_0(k)
  \label{eqn:zeft_pk2}
\end{eqnarray}
where the first term is of the form derived in \cite{TayHam96}, which can
be approximated as an exponential damping of $P_0(k)$ \cite{Mat08a,CLPT}.
The width of the damping term, $\Sigma$, is the 1D rms displacement
computed in linear theory.
We give two approximate forms of the power spectrum, which are equal up
to terms higher order in $P_0$.

\section{Comparison to simulations}
\label{sec:sims}

We shall now compare the models described above to the clustering of dark
matter measured from a series of N-body simulations.
We use 10 simulations run with the TreePM code of \cite{TreePM}, each of the
same ($\Lambda$CDM) cosmology but with a different random number seed chosen
for the initial conditions.  These simulations have been described in more
detail elsewhere \cite{Rei14,RSDmock}, but briefly they were performed in
boxes of size $1380\,h^{-1}$Mpc with $2048^3$ particles and modeled a
$\Lambda$CDM cosmology with $\Omega_m=0.292$, $h=0.69$, $n_s=0.965$
and $\sigma_8=0.82$.
We use outputs at $z=0.25$, $0.50$, $0.75$ and $1.00$ to sample the range of
most interest for upcoming large-scale structure surveys.
For this cosmology $\Sigma=5.3\,h^{-1}$Mpc at $z=0.25$, dropping to
$4.6\,h^{-1}$Mpc by $z=0.5$, $4.1\,h^{-1}$Mpc by $z=0.75$ and $3.7\,h^{-1}$Mpc
by $z=1$.

\begin{figure}
\begin{center}
\resizebox{1.08\columnwidth}{!}{\includegraphics[trim={1.0cm 0 0 0},clip]{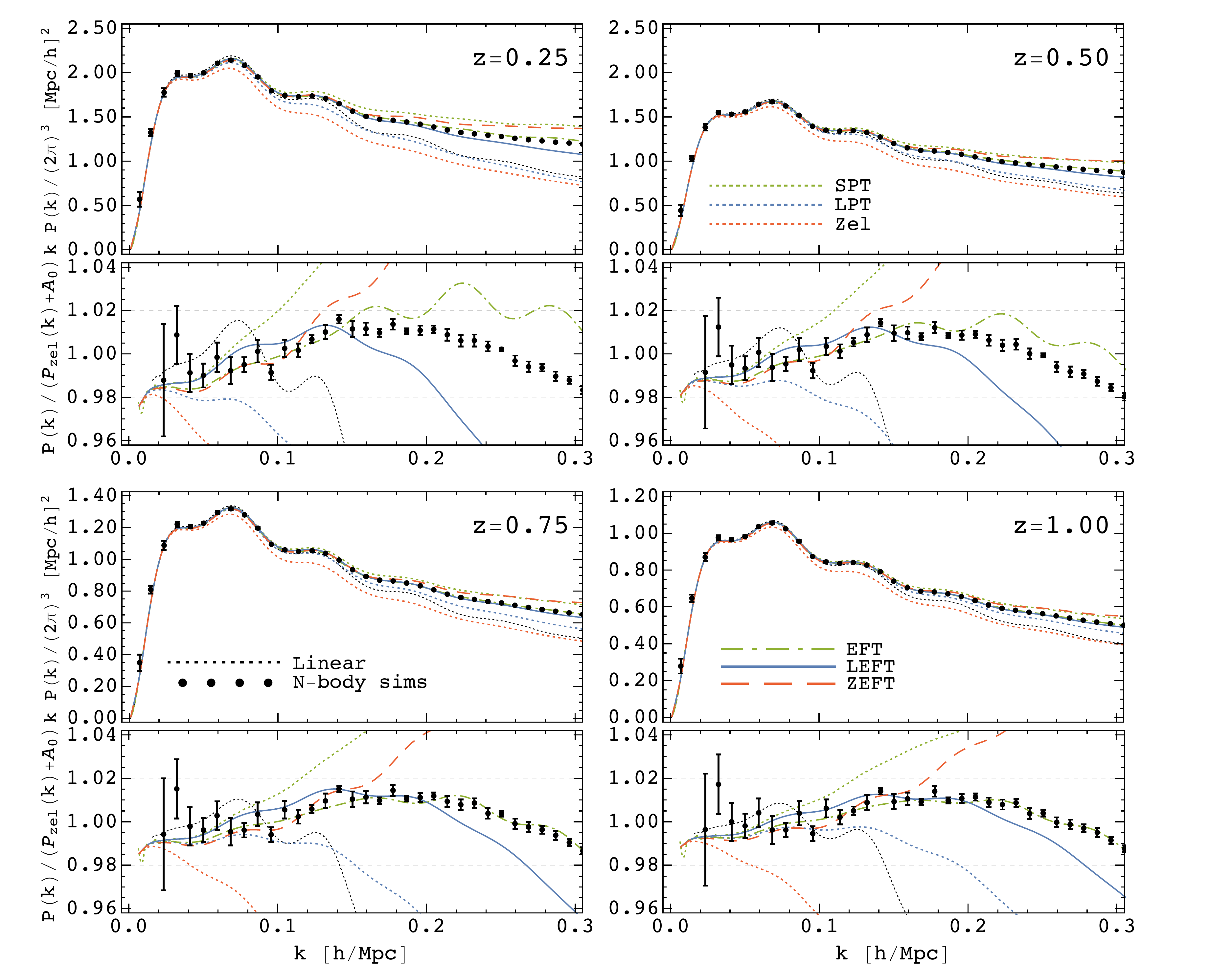}}
\end{center}
\caption{A comparison of the analytic models with N-body simulations at
$z=0.25$ (top left), $z=0.5$ (top right), $z=0.75$ (bottom left) and
$z=1$ (bottom right).
In each panel the black points with error bars represent the mass power
spectrum measured from the simulations described in the text.
The lines show the different analytical models, described previously and
the dotted lines of each color show the perturbation theory results without
the EFT terms.
The upper panels show $k\,P(k)$ while the lower panels, with a zoomed-in
$y$-axis, show the ratio to a fiducial model (taken to be the Zeldovich power
spectrum plus a constant, adjusted as in \cite{SelVla15}).}
\label{fig:ps}
\end{figure}

Each of the EFT models has at least one free parameter, which we adjust by
eye to find the best fit.  Since we have this additional freedom we expect
the EFT models to fit the N-body data better than the `standard' perturbation
theories.
We fix the parameter in Fourier space and use then the same value in
configuration space.
We compare the clustering with both an absolute scale (chosen to be
$k\,P(k)$ in Fourier space and $r^2\,\xi(r)$ in configuration space)
and as a ratio to a fiducial model to enable a better view of the level
of agreement.
We have chosen as our fiducial model a simplified version of the
``Zeldovich halo model'' of ref.~\cite{SelVla15}.
The fiducial model consists of the Zeldovich power spectrum
(section \ref{sec:zeft}) plus a constant,  
$A_0=395$, 240, 145, $92\,h^{-3}{\rm Mpc}^3$ at $z=0.25$, 0.5, 0.75 and 1.
The correlation function of the fiducial model matches the Zeldovich
correlation function except at zero lag, which we shall ignore.

Figure \ref{fig:ps} compares the Fourier space clustering predicted by the
models to that measured in our N-body simulations at various redshifts,
i.e.~the real-space, mass power spectra.
For each redshift the upper panels show the absolute clustering (as
$k\,P(k)$) and the lower panels the ratio to our fiducial model.
In each panel we have fit the amplitude of the EFT terms by eye to give
agreement at the percent level or better at low $k$ while improving the
agreement at intermediate scales.  Different choices of these parameters
can extend or decrease the range of agreement at higher $k$ at the expense
of better or worse agreement at intermediate scales, so this comparison
depends upon the chosen metric (and thus is subjective).
This is particularly true for the Eulerian EFT lines, where much better
agreement at high $k$ can be obtained with a slightly worse fit at low $k$.
Since our N-body simulations are noisy at low $k$ we have chosen to
de-emphasize those points when fitting the EFT model.
We note that all of the models fare quite well for scales up to half of
the non-linear scale
($k_{\rm nl}\equiv\Sigma^{-1}\sim 0.20\,h\,{\rm Mpc}^{-1}$ at $z=0.25$,
$0.22$ at $z=0.5$, $0.24$ at $z=0.75$ and $k=0.27\,h\,{\rm Mpc}^{-1}$ at $z=1$)
and all of the perturbation schemes perform significantly better than
linear theory (black dotted line).
The Lagrangian EFT appears to match the matter clustering as well as
the Eulerian EFT, with the same number of free parameters, at lower $z$ but
the Eulerian formulation outperforms the Lagrangian one at higher $z$.  Since
we know that expanding the exponential in CLEFT would reproduce Eulerian EFT
(see section \ref{sec:comparisonEFT}) this suggests the exponentiation of
$A_{\rm lin}$ is overdamping the power at high $k$.
Eulerian EFT tends to overshoot the N-body points at high $k$ while
Lagrangian EFT undershoots.
In EFT the $k^2\,P(k)$ term comes in with a negative sign, reducing the
well-known overshoot of 1-loop perturbation theory so that it better
matches the N-body result (the standard perturbation theory result is
shown as the dotted green line).  At high $k$ the power goes negative.
The addition of the $k^2\,P(k)$ term with a positive coefficient to the
Zeldovich power spectrum to get the ZEFT model corrects the well-known
short-fall of power in the Zeldovich approximation at high $k$
(the Zeldovich approximation without the $k^2\,P(k)$ term is shown as the
dotted line in the lower panels for each redshift).
In fact, it appears this addition leads to an overestimate of power at high
$k$.  The result is very similar whether we use the approximate forms
eq.~\ref{eqn:zeft_pk1} or \ref{eqn:zeft_pk2} and it persists even at high $z$.
Such a correction could obviously be tamed by including higher-order terms
of the EFT expansion at the cost of introducing further parameters.
Figure \ref{fig:ps} shows that the 1-loop terms present in the CLEFT but not
ZEFT significantly improve the agreement with simulations in the quasi-linear
regime because they give more high $k$ power than the Zeldovich approximation
and thus require less of the $k^2\,P(k)$ term.
At all redshifts the 1-loop CLEFT model agrees with the N-body simulations
better than ZEFT, but the agrement is limited to scales sufficiently below
$k_{\rm nl}$.
It is interesting that the fiducial model manages to agree with the
simulations to nearly as high $k$ as any of the EFTs we have presented,
even though it also has only one parameter which could be tuned.
We shall return to this comparison later.

\begin{figure}
\begin{center}
\resizebox{1.16\columnwidth}{!}{\includegraphics[trim={2.5cm 0 0 0},clip]{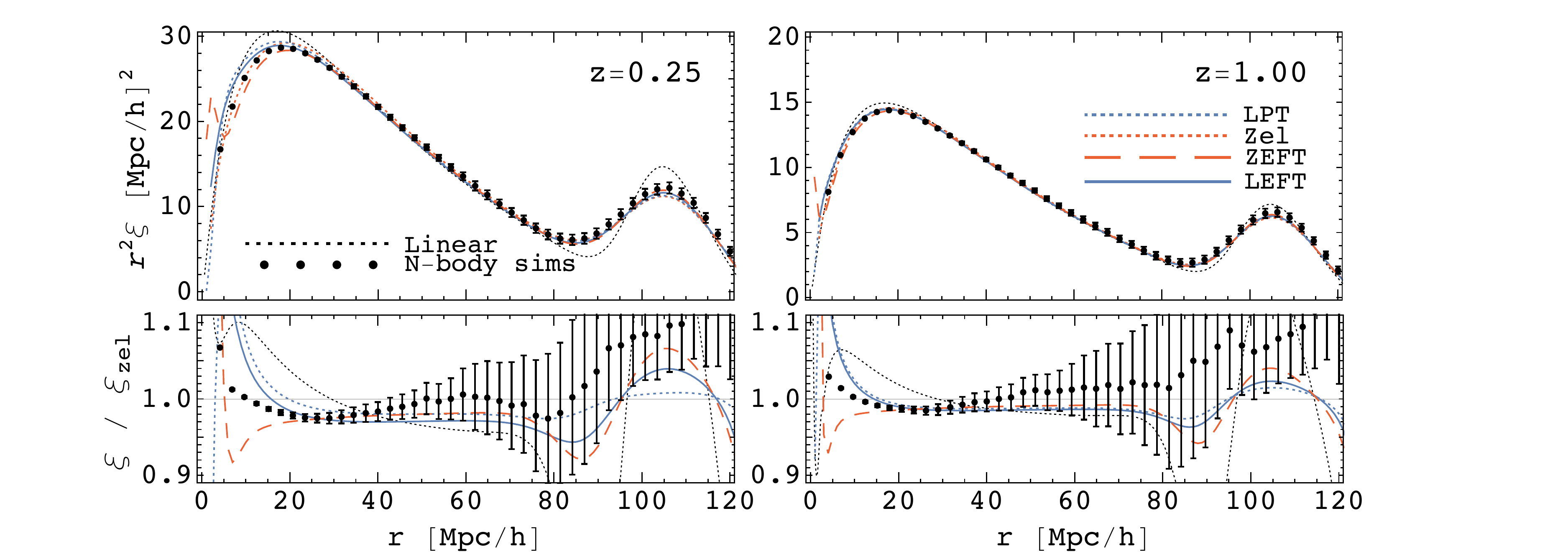}}
\end{center}
\caption{As for figure \ref{fig:ps} but in configuration space and with only
the $z=0.25$ and $z=1$ results.  The upper panels show $r^2\,\xi(r)$ while
the lower panels show the ratio to the fiducial model, which is simply the
Zeldovich approximation (which is thus not plotted).  The error bars, obtained
from the scatter between the 10 N-body simulations, are highly correlated.}
\label{fig:xi}
\end{figure}

Figure \ref{fig:xi} makes the same comparison in configuration space,
i.e.~with the real-space, mass correlation function.
Again the boxes show the different redshifts with the upper panels showing
the absolute clustering and the lower panels the ratio to our fiducial model
(which in this case is simply the Zeldovich approximation).
The N-body data are shown as the points, which have significant and highly
correlated error bars, while the lines show the same approximations as above.
Here we present only the two end cases, $z=0.25$ and $z=1$, to save space.
The Eulerian theories do not make a meaningful prediction for the correlation
function, since the power spectrum diverges as $k\to\infty$.
The correlation function in the Lagrangian theories is, however, quite well
behaved.
As before all of the models fare quite well on scales a factor of $\sim 2-3$
above the non-linear scale, $\Sigma$, and all of the perturbation schemes
perform significantly better than linear theory which misses important
contributions even at $100\,h^{-1}$Mpc.
We note that the Zeldovich approximation does very well on large scales,
as has been noted previously \cite{Tas14a,Whi14}.
The CLEFT model is not a significant improvement over ZEFT in configuration
space, despite doing much better in Fourier space.
The fact that the Zeldovich approximation and ZEFT do so much better in
configuration space than in Fourier space indicates that the Fourier space
error must transform to small lags in configuration space
(see refs.~\cite{Whi14,SelVla15} for further discussion).

\section{Conclusions}
\label{sec:conclusions}

Much of the development of perturbation theory in cosmology has focused on
an Eulerian description in Fourier space.  In this paper we have continued
the development of a Lagrangian description, comparing the performance in
both Fourier and configuration space.  We have extended the traditional
perturbative approach using ``effective field theory'', which amounts to
adding to the solution additional terms that are consistent with the
symmetries of the problem, are arranged in an expansion in derivatives
(powers of $k$), and have free coefficients which must be adjusted to fit
the data.

Throughout we have tried to make connection with earlier work on this
subject, especially refs.~\cite{PorSenZal14,Bal15}.
We provide a slightly different route to the EFT corrections but find the
same functional form as these earlier works.
Specifically, small-scale physics which is neglected within the usual
perturbative calculation gives rise to an additive contribution to the
displacement field.
The lowest order (in derivatives and powers of $\delta_0$) contribution
to $\vPsi$ (which transforms as a vector) is proportional to $\nabla\delta_0$,
with an undetermined coefficient.  There could also be an (unconstrained)
`stochastic' term, which is assumed to be small on large scales.
If we work at the lowest order, these additional terms modify only the
2-point function of the Lagrangian displacement, adding terms proportional
to moments of the linear theory correlation function or closely related
functions.
These terms serve both to tame the dependence of the perturbation theory on
small-scale physics and to improve the agreement with N-body simulations.
We show explicitly that 1-loop, Lagrangian EFT is identical to 1-loop,
Eulerian EFT if the Lagrangian solution is expanded in powers of the linear
theory power spectrum, $P_0$.  However, the Lagrangian formalism also
lends itself to several natural resummation schemes, one of which we adopt
here because it is straightforward both algebraically and numerically
(section \ref{sec:resum}).
We make explicit comparison with the IR-resummation scheme of
ref.~\cite{SenZal15}, showing how to derive their formulae within our
framework in appendix \ref{app:IRresum}.
We choose to resum all of the linear pieces of $A_{ij}$, rather than just
the low $k$ terms as in \cite{McQWhi15}.  This leads to little change in
the final results.

Comparing the model predictions to the 2-point clustering of the mass in
N-body simulations in both Fourier and configuration space, we find that
all of the EFT models fare quite well on scales larger than about twice the
non-linear scale, but systematically fail as the non-linear scale is
approached.
All of the perturbative schemes fare better than linear theory on
quasi-linear scales.
In Fourier space the Lagrangian EFT matches the N-body simulations as well as
the Eulerian EFT at low redshifts, but fails at larger scales at high $z$.
Since we know that Lagrangian EFT reproduces Eulerian EFT if we expand to a
fixed order in the linear power spectrum (section \ref{sec:comparisonEFT}) this
suggests than our resummation scheme is overdamping power at large $k$.
Adding the $k^2\,P(k)$ contribution to the Zeldovich approximation partially
corrects the well-known short-fall of power at high $k$ for this model, but
leads to an over-prediction of power on small scales.
Including the 1-loop contributions significantly improves the agreement with
the N-body results on quasi-linear scales.
The Eulerian EFT does not make a finite prediction for the correlation
function, but the Lagrangian form does and the agreement with the simulations
is very good on large scales.
In fact we find that all of the Lagrangian schemes are in very good agreement
with the N-body simulations: including the 1-loop corrections, going from ZEFT
to CLEFT, leads to only a modest improvement in the agreement with simulations.
The fact that ZEFT does so much better in configuration space than in Fourier
space indicates that the Fourier space error must transform to small lags in
configuration space, i.e.~it must consist primarily of `broad band' power.

In this paper we have developed and tested a Lagrangian effective field
theory to predict the low order clustering of the matter field in real space
(addressing many of the same themes as ref.~\cite{PorSenZal14}).
A {\tt C++} code to compute the formulae presented above given a linear
theory power spectrum is publicly available at
{\tt https://github.com/alejandroaviles/CLEFT}.
One of the advantages of the Lagrangian approach is the relative ease with
which redshift-space distortions and bias can be incorporated.
We plan to consider these developments in a future paper.

\acknowledgments
We would like to thank Tobias Baldauf, Matt McQuinn, Uros Seljak 
and Matias Zaldarriaga for useful discussions and for comments on an
early draft of the manuscript.

Z.V.~is supported in part by the U.S. Department of Energy contract to
SLAC no.~DE-AC02-76SF00515.
A.A.~is supported by the UCMEXUS-CONACyT Postdoctoral Fellowship.

The analysis in this paper made use of the computing resources of the
National Energy Research Scientific Computing Center.

\appendix

\section{Useful identities}
\label{app:identities}

To simplify the mode coupling integrals we follow \cite{Mat08b} and define
\begin{equation}
  R_n(k) \equiv \frac{k^3}{(2\pi)^2} P_0(k) \int_0^\infty dr\ P_0(kr)
  \widetilde{R}_n(r)
\label{eqn:Rndef}
\end{equation}
with
\begin{equation}
  \widetilde{R}_1=\int_{-1}^{1}d\mu\ \frac{r^2(1-\mu^2)^2}{1+r^2-2r\mu}
  \quad , \quad
  \widetilde{R}_2=\int_{-1}^{1}d\mu\ \frac{(1-\mu^2)r\mu(1-r\mu)}{1+r^2-2r\mu}
\end{equation}
we also use
\begin{equation}
  Q_n(k) \equiv \frac{k^3}{(2\pi)^2} \int_0^\infty dr\ P_0(kr)
  \int_{-1}^{+1}d\mu\ P_0\left( k\sqrt{1+r^2-2r\mu} \right)
  \widetilde{Q}_n(r,\mu)
\label{eqn:Qtdef}
\end{equation}
with
\begin{eqnarray}
  \widetilde{Q}_1 &=& \frac{r^2(1-\mu^2)^2}{(1+r^2-2r\mu)^2} \\
  \widetilde{Q}_2 &=& \frac{(1-\mu^2)r\mu(1-r\mu)}{(1+r^2-2r\mu)^2} \\
  \widetilde{Q}_3 &=& \frac{\mu^2(1-r\mu)^2}{(1+r^2-2r\mu)^2}
\end{eqnarray}
In the limit of high-$k$ we have $R_1(k)\to (8/5)(k^2\Sigma^2)P_0(k)$ and
$R_2(k)\to (-2/5)(k^2\Sigma^2)P_0(k)$.

In performing these calculations one frequently needs to do Gaussian
integrals.  Again to simplify our notation, and make connection with
earlier work, we define
\begin{equation}
  G(\vk) = e^{-\frac{1}{2}k_ik_jA_{ij}+ib_ik_i}
  \quad , \quad
  G(\vec{b}) = \frac{1}{(2\pi)^{3/2}|\vec{A}|^{1/2}}
  e^{-\frac{1}{2}b_ib_jA_{ij}^{-1}}
\end{equation}
in terms of which
\begin{eqnarray}
\int\frac{d^3k}{(2\pi)^3} G(\vk)         &=&              Q(\vec{b})\\
\int\frac{d^3k}{(2\pi)^3} G(\vk)k_a      &=& ig_a         Q(\vec{b})\\
\int\frac{d^3k}{(2\pi)^3} G(\vk)k_ak_b   &=& G_{ab}       Q(\vec{b})\\
\int\frac{d^3k}{(2\pi)^3} G(\vk)k_ak_bk_c&=& i\Gamma_{abc}Q(\vec{b})
\end{eqnarray}
with
$\Gamma_{ijk}=
 \left(\vec{A}^{-1}\right)_{ij}g_k+
 \left(\vec{A}^{-1}\right)_{ki}g_j+
 \left(\vec{A}^{-1}\right)_{jk}g_i-g_ig_jg_k$,
$G_{ij}=\left[\left(\vec{A}^{-1}\right)_{ij}-g_i g_j\right]$ and
$\vec{g}=\vec{A}^{-1}\vec{b}$ using the notation of
Ref.~\cite{CLPT}.

Finally the following integrals are useful when evaluating the power
spectrum, as they allow the integral over $d^3q$ to be reduced to a
sum of 1D integrals:
\begin{eqnarray}
  \int d\mu\ e^{i\mu\,A+\mu^2\,B} &=& 2e^B\sum_{\ell=0}^\infty
  \left(-\frac{2B}{A}\right)^\ell j_\ell(A) \\
  \int d\mu\ \mu\,e^{i\mu\,A+\mu^2\,B} &=& 2ie^B\sum_{\ell=0}^\infty
  \left(-\frac{2B}{A}\right)^\ell j_{\ell+1}(A) \\
  \int d\mu\ \mu^2\,e^{i\mu\,A+\mu^2\,B} &=& 2e^B\sum_{\ell=0}^\infty
  \left(1+\frac{\ell}{B}\right)\left(-\frac{2B}{A}\right)^\ell j_\ell(A) \\
  \int d\mu\ \mu^3\,e^{i\mu\,A+\mu^2\,B} &=& 2ie^B\sum_{\ell=0}^\infty
  \left(1+\frac{\ell}{B}\right)\left(-\frac{2B}{A}\right)^\ell j_{\ell+1}(A)
\end{eqnarray}

\section{IR resummation}
\label{app:IRresum}

Senatore and Zaldarriaga have put forth an IR-resummation scheme for EFT
\cite{SenZal15} which appears different from the schemes we discussed in the
main text.  In this appendix we show how to derive their scheme from ours.
For simplicity we shall restrict ourselves to the Zeldovich approximation,
thus $A_{ij}$, $\Sigma^2$ etc.~are to be evaluated in linear theory.
This already captures all of the main features of the scheme (and indeed most
of the main physical effects \cite{Tas14a,Whi14}) and the generalization to
higher orders is straightforward.

We will focus on the power spectrum, which we can write following
eq.~(\ref{eqn:pofk}) as
\begin{eqnarray}
  P(k) &=& \int d^3q\ e^{i\vk\cdot\vq}
           \left(e^{-(1/2)k_ik_jA_{ij}}-1\right) \\
  &=& \int d^3q\ e^{i\vk\cdot\vq}\ e^{-k^2\Sigma^2}
      \left(e^{k_ik_j\eta_{ij}}-1\right) \\
  &=& \int d^3q\ e^{i\vk\cdot\vq}\ e^{-k^2\Sigma^2+k_ik_j\eta_{ij}}
      \left( 1-e^{-k_ik_j\eta_{ij}} \right) \\
  &=& \int d^3q\ e^{i\vk\cdot\vq}\ e^{-k^2\Sigma^2+k_ik_j\eta_{ij}}
      \left( k_ik_j\eta_{ij}-\frac{1}{2}\left[k_ik_j\eta_{ij}\right]^2
      +\cdots \right) \\
  &=& \int d^3q\ e^{i\vk\cdot\vq}\ e^{-k^2\Sigma^2+k_ik_j\eta_{ij}}
      \left[k_ik_j\eta_{ij}\right]
      \left(1+\left[k^2\Sigma^2-k_ik_j\eta_{ij}\right]\right) \nonumber \\
  &+& \int d^3q\ e^{i\vk\cdot\vq}\ e^{-k^2\Sigma^2+k_ik_j\eta_{ij}}
      \left(\frac{1}{2}\left[k_ik_j\eta_{ij}\right]^2-k^2\Sigma^2
      \left[k_ik_j\eta_{ij}\right]\right)  + \cdots
\end{eqnarray}
where we remind the reader that $A_{ij}$ and $\eta_{ij}$ are functions of
the Lagrangian displacement, $\vq$.
If we now define
$\mathcal{K}(k,q)=\exp\left[-k^2\Sigma^2+k_ik_j\eta_{ij}\right]$
we have
\begin{eqnarray}
  \mathcal{K}^{-1}(k,q) &=& \exp\left[k^2\Sigma^2-k_ik_j\eta_{ij}\right] \\
  \left. \mathcal{K}^{-1}(k,q) \right|_N &=&
  \sum_{n=0}^N\frac{(k^2\Sigma^2-k_ik_j\eta_{ij})^n}{n!}
\end{eqnarray}
adopting the notation of \cite{SenZal15}.
It is useful to note that $k_ik_j\eta_{ij}$ Fourier transforms to $P_0(k)$.
Thus
\begin{eqnarray}
  P(k) &=& \int d^3q\ e^{i\vk\cdot\vq}\ \mathcal{K}(k,q)
  \left. \mathcal{K}^{-1}(k,q)\right|_1 \left[k_ik_j\eta_{ij}\right]
  \nonumber \\
  &+& \int d^3q\ e^{i\vk\cdot\vq}\ \mathcal{K}(k,q) \left(
  \frac{1}{2}\left[k_ik_j\eta_{ij}\right]^2-
  k^2\Sigma^2\left[k_ik_j\eta_{ij}\right]\right) + \cdots
\end{eqnarray}
If carried through to infinite order this resummation scheme would
reproduce the Zeldovich approximation (which our resummation scheme
naturally does), but at any finite order some terms are missed.

Our scheme naturally sums all of the linear terms, which has proven to
be practically useful to match the broadening of the BAO feature in the
correlation function.  If we wish to preserve the translation invariance
of the theory, there are two obvious ways to do such a resummation.
One is to do a `linearization', as in ref.~\cite{SenZal15}, which includes
terms based on their perturbation theory order.
The other is the one we have advocated above.

\bibliographystyle{JHEP}
\bibliography{ms}
\end{document}